\documentclass[twocolumn,showpacs,preprintnumbers,amsmath,amssymb,aps,prb]{revtex4}

\pdfoutput=1

\usepackage{graphicx}
\usepackage{dcolumn}
\usepackage{bm}
\usepackage{subfigure}
\usepackage{natbib}
\usepackage{multirow}

\newcommand{\oneA}{\,\mathrm{\AA}^{-1}}

\begin{document}

\title{Direct observation of electron thermalization and electron-phonon coupling in photoexcited bismuth \\}

\author{J. Faure$^{1}$, J. Mauchain$^{2}$, E. Papalazarou$^{2}$, M. Marsi$^{2}$, D. Boschetto$^{1}$, I. Timrov$^{3}$, N. Vast$^{3}$, Y. Ohtsubo$^{4}$, B. Arnaud$^{5}$, and L. Perfetti$^{3}$}

\affiliation{
$^{1}$ Laboratoire d'Optique Appliqu\'{e}e, Ecole Polytechnique-ENSTA-CNRS UMR 7639, 91761 Palaiseau, France}

\affiliation{
$^{2}$ Laboratoire de Physique des Solides, CNRS-UMR 8502, Universit\'e Paris-Sud, F-91405 Orsay, France}

\affiliation{
$^{3}$ Laboratoire des Solides Irradi\'{e}s, Ecole Polytechnique - CEA/DSM - CNRS UMR 7642,  91128 Palaiseau, France}

\affiliation{
$^{4}$ Synchrotron SOLEIL, Saint-Aubin-BP 48, F-91192 Gif sur Yvette, France}

\affiliation{
$^{5}$ Institut de Physique de Rennes (IPR), UMR UR1-CNRS 6251, F-35042 Rennes Cedex, France}

\date{\today}

\begin{abstract}
We investigate the ultrafast response of the bismuth (111) surface by means of time resolved photoemission spectroscopy. The direct visualization of the electronic structure allows us to gain insights on electron-electron and electron-phonon interaction. Concerning electron-electron interaction, it is found that electron thermalization is fluence dependent and can take as much as several hundreds of femtoseconds at low fluences. This behavior is in qualitative agreement with Landau's theory of Fermi liquids but the data show deviations from the behavior of a common 3D degenerate electron gas. Concerning electron-phonon interaction, our data allows us to directly observe the coupling of individual Bloch state to the coherent $A_{1g}$ mode. It is found that surface states are much less coupled to this mode when compared to bulk states. This is confirmed by \textit{ab initio} calculations of surface and bulk bismuth.

\end{abstract}


\maketitle

\section{\label{sec:Introduction}Introduction}
With the development of ultrafast pump-probe experiments in solids, it is now possible to create strongly out of equilibrium states of matter using a femtosecond pump laser pulse. Exotic ultrafast phenomena such as non thermal melting, \cite{rous01,scia09} or excitations of coherent phonons \cite{zeig92,hase98,soko03} have been observed in the past decade. Other than exotic states of matter and ultrafast phase transitions, this experimental technique is also important because it can give new insights on fundamental interactions which occur in a solid: interactions occurring on different timescales become accessible through temporal discrimination. The classical picture used to describe the effect of photoexcitation by a femtosecond laser pulse in a metal is the following (i) electrons are first promoted to higher energy states through direct optical transitions (ii) electrons thermalize through electron-electron interaction (iii) the electron bath and the lattice then thermalize on a time-scale given by the electron-phonon interaction. In addition to these population dynamics, the ultrafast light pulse can also generate a macroscopic polarization of electrons and the nuclear lattice. Within the displacive limit, photoexcitation changes the shape of the potential energy surface with respect to the lattice coordinates. The transient shift of energy minima can drive coherent oscillations of atomic positions, i.e. a coherent phonon. This is the displacive excitation of coherent phonon (DECP) mechanism.\cite{zeig92,fritz07}

Bismuth has been intensely studied in this context because it is a good testbed material for pump-probe experiments. Bismuth is subject to a Peierls distortion which breaks the translational symmetry along the (111) direction.\cite{hofm06} This makes this material very sensitive to the electronic distribution as small changes in the electron occupation can affect the potential energy surface and trigger atomic motion in the $A_{1g}$ mode. Another peculiar feature of bismuth is that the density of states close to the Fermi level is very small. This, of course, explains why bismuth is a semi-metal. But more importantly, this restricts the phase space available to scattering events and consequently slows down the dynamics of electron-electron and electron-phonon interactions. Following excitation, it takes a long time for a bismuth sample to return to equilibrium, typically tens of picoseconds, which is why coherent phonon oscillations are well described by the displacive mechanism.\cite{zeig92,gire11}

In this article, we take advantage of this relatively slow dynamics in bismuth to explore electron-electron interaction and the time it takes for the electron bath to thermalize. This aspect is usually not studied experimentally as most experiments use optical pump-probe methods \cite{hase98,bosc08} as well as time-resolved diffraction techniques,\cite{soko03,fritz07,john08,scia09} which do not permit a direct visualization of the electronic states. In contrast, we have used time-resolved and angle resolved photoemission spectroscopy (ARPES) which is the ideal tool to investigate electron thermalization because it permits the direct visualization of electronic states in reciprocal space.\cite{perf06,perf07,graf11,cort11,kirs10,ishi11,rohw11} 

The paper is organized as follows: section \ref{sec_Bi} discusses the static electronic structure of bismuth as observed using ARPES. Section \ref{sec_therm} focuses on electron thermalization which occurs in tens to hundreds of femtoseconds, depending on the fluence of the excitation laser pulse. Section \ref{sec_elph} focuses on the next few picoseconds following photoexcitation and new information concerning the coupling of individual electron states with the $A_{1g}$ phonon mode are revealed. Finally the last section brings in the conclusions.

\section{Electronic structure of the Bi (111) surface}\label{sec_Bi}

The experiment was performed on the FemtoARPES setup, using a Ti:sapphire laser system delivering 35 fs pulses at 780 nm with 250 kHz repetition rate.\cite{faur12} Part of the laser beam is used to generate a UV probe pulse for photoemission. The 6.3 eV photons are obtained through cascade frequency mixing in BBO crystals. This UV source permits to perform time-resolved photoemission experiments with an energy resolution better than 70 meV (limited by the energy bandwidth of the UV pulses) and a temporal resolution better than 65 fs.\cite{faur12} The probe pulse is focused to about $50\,\mu m$  at Full Width Half Maximum (FWHM) on the sample whereas the 1.6 eV pump pulse is focused to about $100\,\mu m$ FWHM. The (111) surface of bismuth was initially optically polished and is prepared under ultra-high vacuum by sputtering and annealing cycles. All experiments are performed at 130 K base temperature and base pressure $<10^{-10}$ mbars.

\begin{figure}[t]
\begin{center}
\includegraphics[width=8.5 cm]{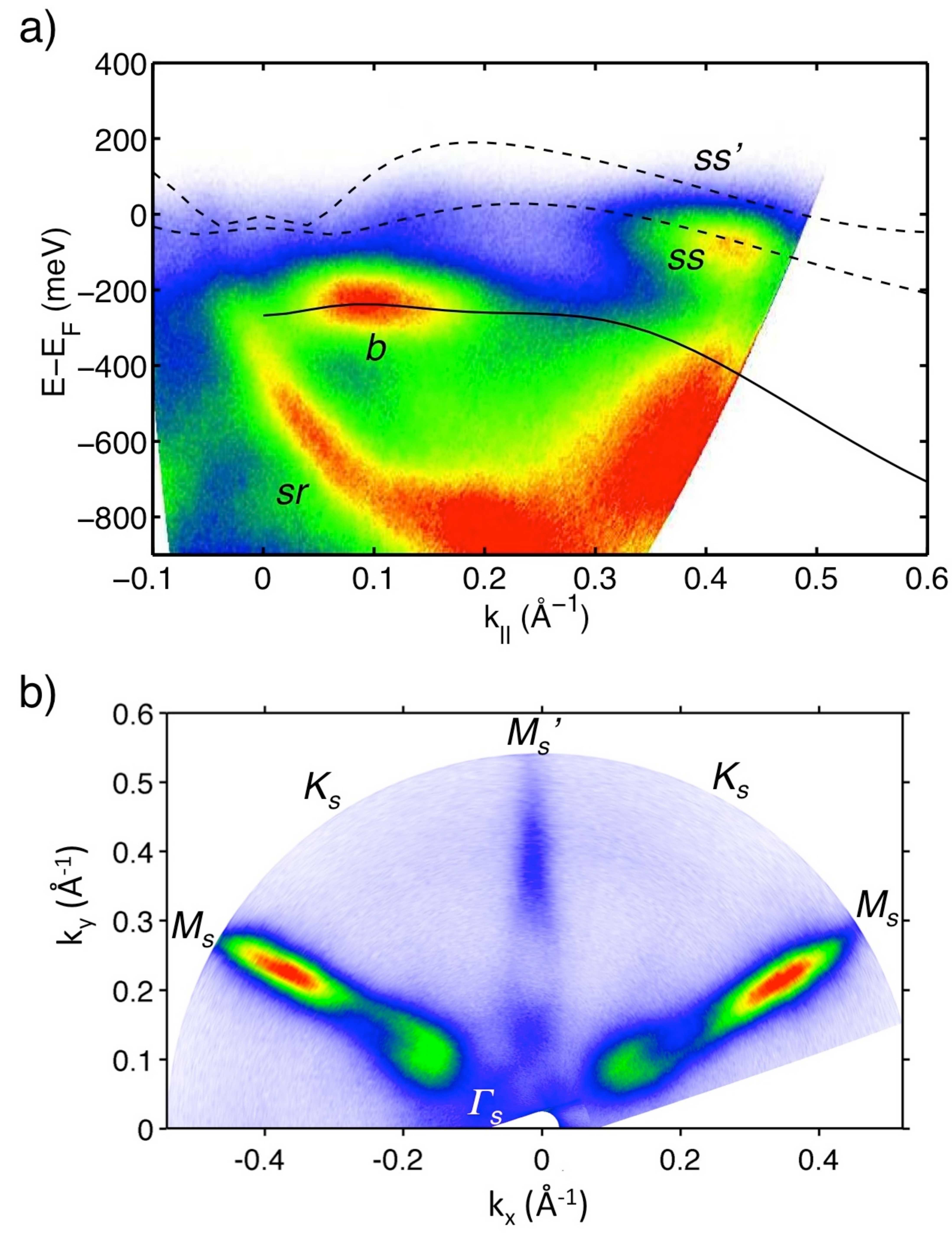}
\caption{a) Angle resolved photoemission data taken on Bi (111) surface at 130 K and along the $\Gamma_s$-$\mathrm{M}_s$ direction of the surface Brillouin zone. \textit{b} is a bulk band whereas \textit{ss} is a surface band and \textit{sr} a surface resonance. Solid line: DFT-LDA calculation of bulk band $b$; dashed line: DFT-GGA calculation of surface state $ss$. (b) Fermi surface of the Bi (111) surface.}
\label{fig_rawdata}
\end{center}
\end{figure}

Figure \ref{fig_rawdata}a shows an angle resolved map of photoelectrons emitted along the $\Gamma_s$-$\mathrm{M}_s$ direction of the surface Brillouin zone. One can observe complex and well defined structures forming various electronic bands. Photoemission at 6.3 eV provides both surface and bulk sensitivity, thus it is necessary to disentangle the data of fig. \ref{fig_rawdata}a in order to understand the bulk/surface character of the various bands. This was done extensively in previous work using symmetry considerations:\cite{papa12} the structure labeled $b$ is a bulk band whereas $ss$ is a surface state and $sr$ a surface resonance. In the following, we will focus on structure $b$ and $ss$ which provide well defined structures both in energy and wave vector. 

Confirmation of the bulk/surface nature of the bands can be obtained using Density Functional Theory (DFT) calculations.  Calculations of the electronic structure of bulk bismuth were performed in the Local Density Approximation (LDA) with the ABINIT package \cite{gonz09} and taking the spin-orbit coupling interaction (SOC) into account. An additional difficulty arises here: since the photoemission process does not conserve the perpendicular component of the wave vector, simply assigning a band to a peak of the photoemission map is not formally correct. However, structure $b$ could be well fitted by a bulk band with perpendicular wave vector $k_\bot=0.45\oneA$ , see full black line in fig. \ref{fig_rawdata}a. DFT calculations for the surface states were done using the WIEN2K code \cite{blaha90} with an exchange-correlation functional based on the generalized gradient approximation (GGA).\cite{perd96} The surface was modeled by a symmetric slab of 20 Bi layers repeated along the (111) direction with a periodic gap.
Structure $ss$ could be well fitted by a surface band dispersing close to the Fermi level, see the black dashed line on fig. \ref{fig_rawdata}a. This band crosses the Fermi level at 2 various locations along the $\Gamma_s$-$\mathrm{M}_s$ direction and for wave vectors between $0.15\oneA$ and $0.35\oneA$. Another surface band (labeled $ss'$ in the figure) also crosses the Fermi level close to the $\Gamma_s$ point. Thus, these bands forms a large Fermi surface which implies that the surface of bismuth is a good metal, contrary to the bulk, which is known to be a semi-metal. By performing an azimuthal rotation of the sample we were able to measure the Fermi surface of the bismuth surface states, see fig. \ref{fig_rawdata}b. Our data, which was measured with 6.3 eV photons, agrees well with data taken on synchrotron light sources with higher energy photons.\cite{ast01,hofm06} 

We recall that at the surface, inversion symmetry is broken so that the SOC lifts the spin degeneracy of surface states. Indeed, the degeneracy of bands $ss$ and $ss'$ is lifted by SOC so that these bands are spin polarized.\cite{koro04,ohts12} This effect has been observed in simulations as well as in spin-resolved photoemission measurements.\cite{kimu10} Note also that our data reproduces closely one step photoemission simulations performed in Ref. \onlinecite{kimu10}, in which the analysis of the polarization of the bands also confirm our interpretation that $b$ and $ss$ are respectively bulk and surface bands.

\section{Dynamics of electron thermalization}\label{sec_therm}

\subsection{Experimental results}
We now turn to the dynamical aspect of the electronic structure following photoexcitation. Photoexcitation was obtained using a 780 nm, 35 fs pump pulse, with fluence in the range $0.01-1\,\mathrm{mJ/cm^2}$. 

For this study, we have focused on the surface band around $k_\|=0.38\oneA$ for observing electron thermalization. Before discussing these results in details, we would like to point out that at 6.3 eV photon energy, photoemission spectroscopy has some bulk sensitivity, typically on a few nanometers. Thus, we are able to probe the dynamics of bulk states in the vicinity of the surface. Conversely, bulk states located deeper in the sample are not accessible. However, it is reasonable to assume that bulk states in the vicinity of the surface and surface states follow similar thermalization dynamics. In consequence, by observing the dynamics of the surface states, it is reasonable to consider that we actually have access to the thermalization of electrons at the surface, whether they are surface or bulk states. 

In general, electron thermalization is assumed to be extremely fast, typically faster than the pump pulse duration. While this might be true for some metals, it is not necessarily the case for other materials. For instance, using time-resolved THz spectroscopy, it has been observed that electron thermalization in Bi can occur on a 0.6 ps time scale when excitation is performed at low fluences,\cite{iuri2012} e.g. $10\,\mathrm{\mu J/cm}^2$. It has been proposed that electron thermalization is slowed down because electrons and holes accumulate for a while in valleys of the conduction and valence bands respectively. 

\begin{figure}[t]
\begin{center}
\includegraphics[width=8.5 cm]{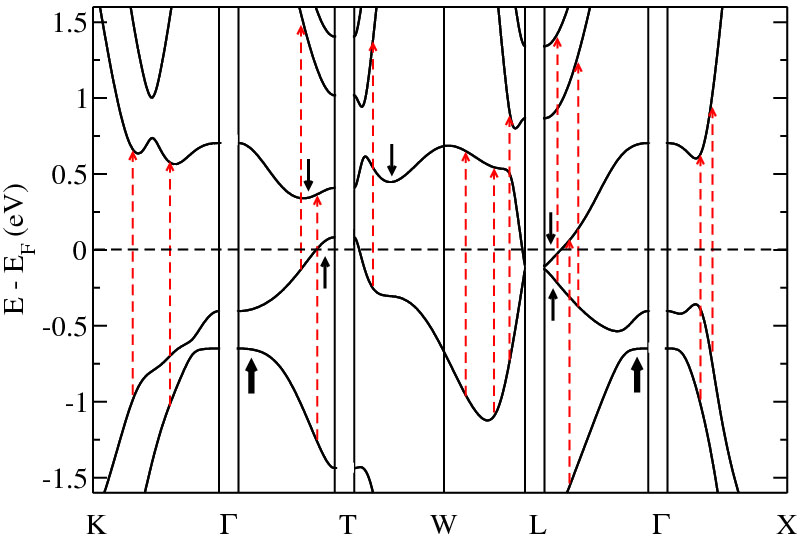}
\caption{DFT-GGA simulations of the electronic structure of bulk Bi, taken from Ref. \onlinecite{iuri2012}. The dashed red arrows stand for direct transitions caused by the 1.6 eV pump laser pulse.}
\label{fig_DFT}
\end{center}
\end{figure}

The arrival of the pump pulse at 1.6 eV promotes electrons from the valence bands to higher energy states in the conduction bands through direct optical transitions. This occurs only in specific locations of the bulk Brillouin zone, as shown by the red arrows in fig. \ref{fig_DFT}. The black arrows in fig. \ref{fig_DFT} indicates true local minima (maxima) in the conduction (valence) bands where electrons (holes) could possibly accumulate, as suggested in Ref. \onlinecite{iuri2012}. These local minima are located along T-W and $\Gamma$-T in the bulk Brillouin zone so that their projections are close to $\Gamma_s$ in the surface Brillouin zone. Thus, we expect to find excited electrons in the 0.5-1 eV region above the Fermi level and close to the $\Gamma_s$ point, as most other possible excitations are located close to the edge of the Brillouin zone which we cannot access with our current set-up (with photon energy of 6.3 eV, one can only measure energy states in Bi up to $k_\|\simeq0.6\,\mathrm{\AA}^{-1}$). DFT-GGA calculations of the surface bands show that direct transitions at 1.6 eV can also occur in the $\Gamma_s$-$\mathrm{K}_s$ direction for surface states. However, we expect most excited electrons to be located in bulk states as the pump laser pulse penetrates more than 15 nm in the bulk, i.e. several bismuth bilayers.

\begin{figure}[t]
\begin{center}
\includegraphics[width=8.5 cm]{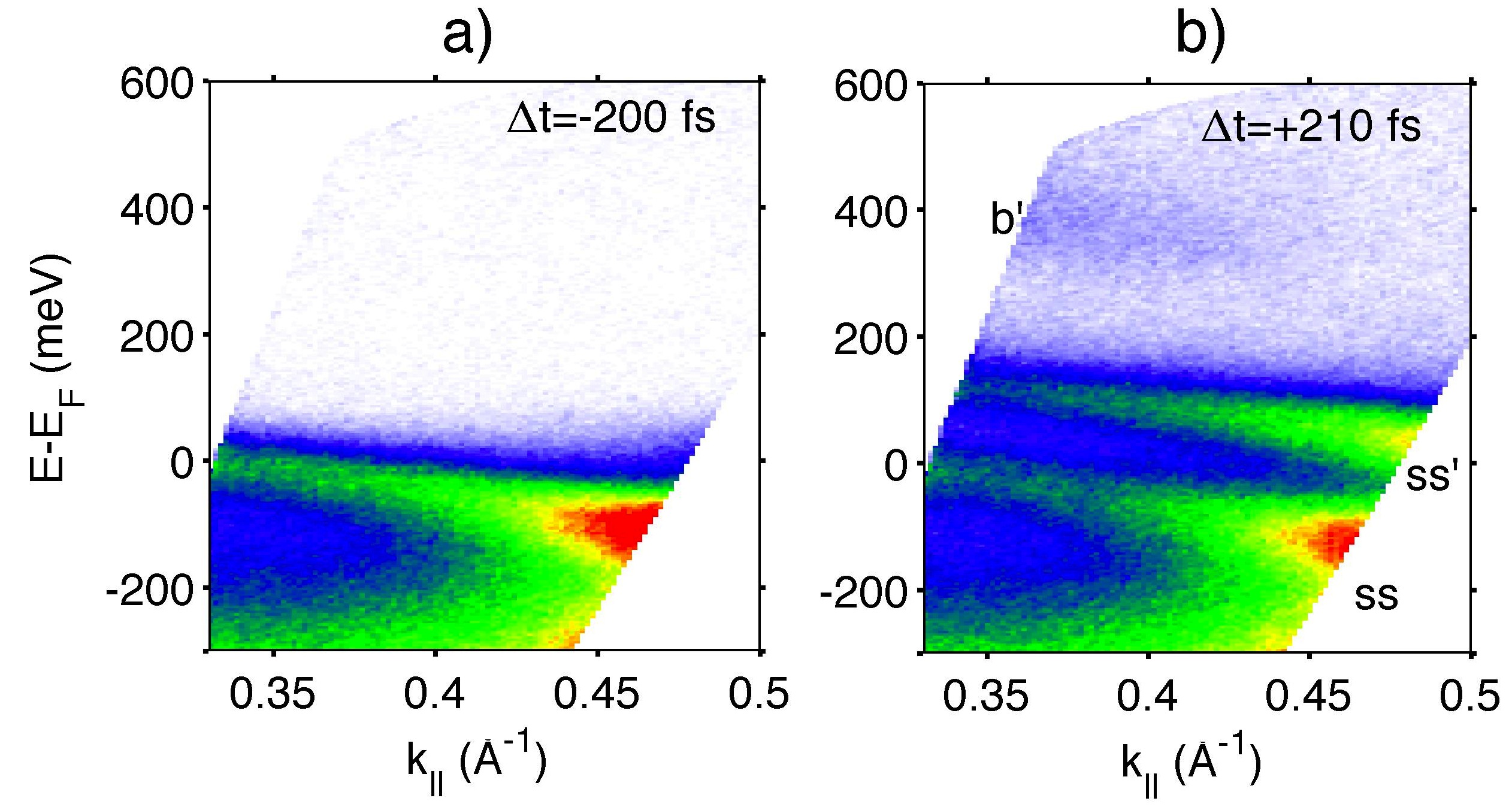}
\caption{Photoelectron intensity map in the $\Gamma_s$-$\mathrm{M}_s$ direction. a) Before the arrival of the pump pulse, b) after the arrival of the pump pulse, fluence $F=0.22\,\mathrm{mJ/cm}^2$. The polarization of the probe pulse is close to  \textit{s}-polarization, which reveals the surface states above and below the Fermi level.}
\label{fig_data}
\end{center}
\end{figure}

In the experiment, we could not locate excited states around the $\Gamma_s$ point although they were predicted theoretically. This might be due to vanishing matrix elements of the dipole operator. Similarly, we could not confirm that electrons accumulate in local minima of the conduction band near $\Gamma_s$, as proposed in Ref. \onlinecite{iuri2012}. However, we were able to detect a large amount of excited electrons close to $k_\|\simeq0.4\,\mathrm{\AA}^{-1}$, i.e. close to the intersection of surface band $ss$ with $E_F$. By using an  \textit{s}-polarized probe beam, we could reveal the structure of the electronic states above the Fermi level, see fig. \ref{fig_data}b. Figure \ref{fig_data}a shows the photoelectron intensity map 200 fs before the pump pulse has arrived; band $ss$, is clearly visible. Figure \ref{fig_data}b shows the situation 215 fs after the pump pulse has arrived: a band above the Fermi level, labeled as $ss'$, has been populated. There is also a small amount of higher energy electrons in the 300-600 meV window, labeled as $b'$. These electronic states at higher energy are likely to be bulk states as they match projected bulk bands from DFT simulations.\cite{ohts12} The two bands close to the Fermi level originally come from the same band which has been split because the SOC lifted the spin degeneracy for the surface states.\cite{koro04} These bands have opposite parity which explains why the band below $E_F$ is clearly visible for \textit{p} polarization whereas the band above $E_F$ is visible for \textit{s} polarization.\cite{ohts12} In the following, the analysis of dark states was carried out using  \textit{s}-polarized light while \textit{p}-polarized light was used for states below $E_F$. 

\begin{figure}[t]
\begin{center}
\includegraphics[width=8.5 cm]{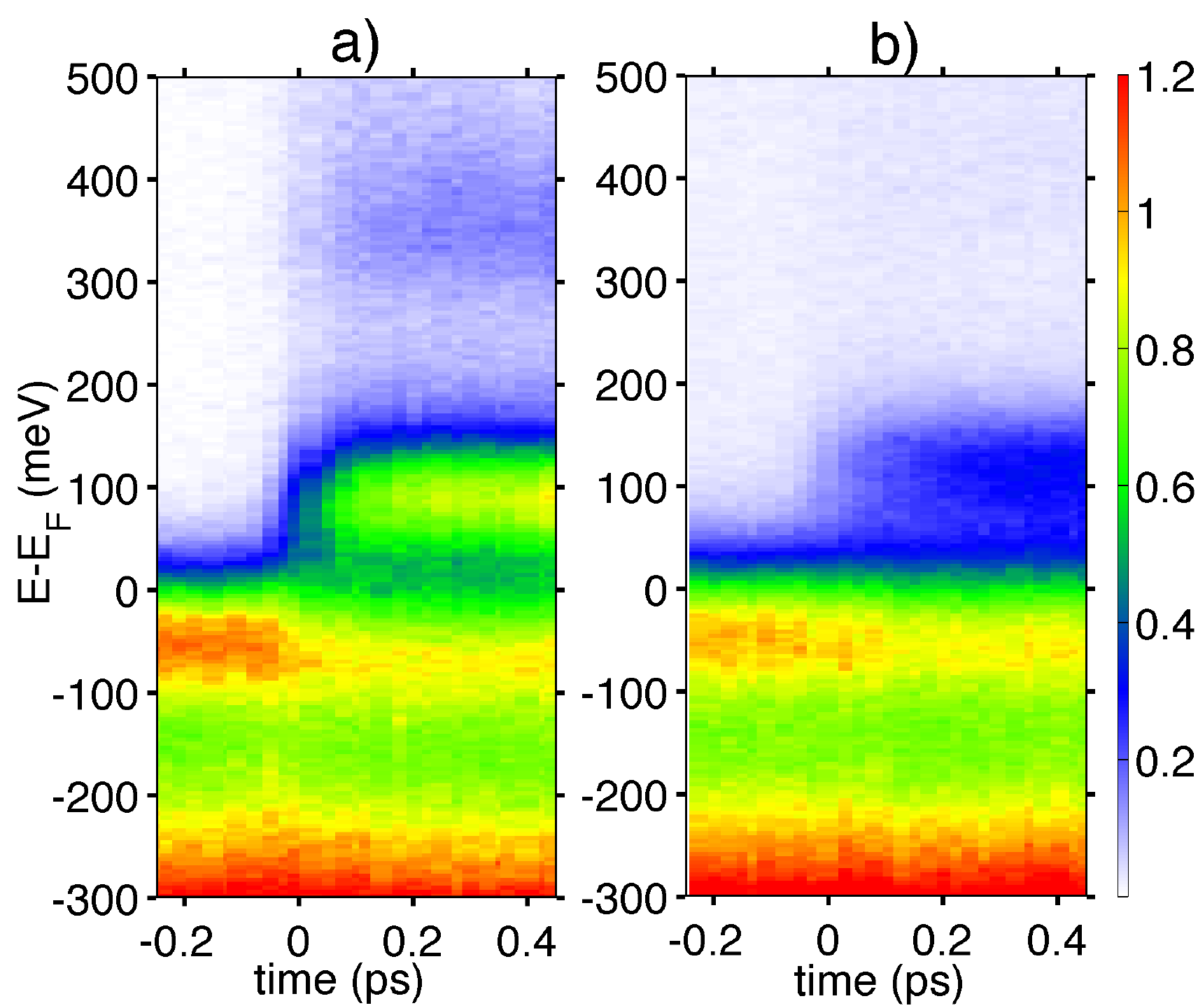}
\caption{Photoelectron spectra at $k_\|=0.38\,\mathrm{\AA^{-1}}$ as a function of pump probe delay. Panel (a) and (b) were taken with a pump fluence of $0.22\,\mathrm{mJ/cm^2}$ and $0.04\,\mathrm{mJ/cm^2}$ respectively.}
\label{fig_spectra}
\end{center}
\end{figure}

\begin{figure}[t]
\begin{center}
\includegraphics[width=8.5 cm]{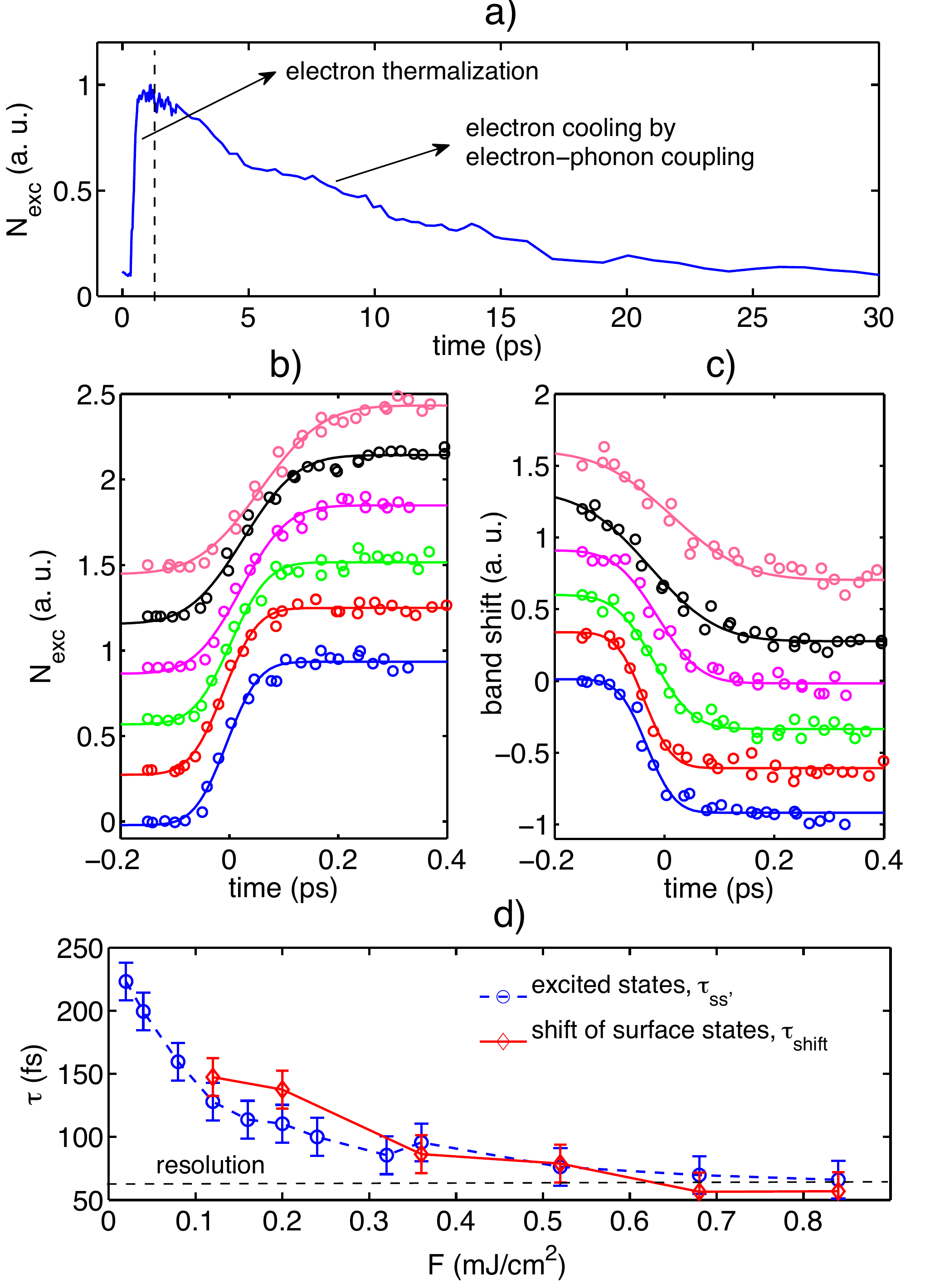}
\caption{(a) Number of excited electrons in state $ss'$ as a function of pump-probe delay for a long scan (fluence was $0.84\;\mathrm{mJ/cm}^2$).
(b) Number of excited electrons in state $ss'$ for different fluences and just after photoexcitation. From pink line to blue line, fluences are $0.12-0.2-0.36-0.52-0.68-0.84\;\mathrm{mJ/cm}^2$. (c) The shift of the surface band as a function of pump-probe delay for the same fluences as in (b). (d) The characteristic time associated with the number of excited states and shift of the surface state \textit{ss} as a function of pumping power.}
\label{fig_dynam}
\end{center}
\end{figure}

Time resolved spectra for $k_\|=0.38\,\mathrm{\AA}^{-1}$ are shown in fig. \ref{fig_spectra} for two different pump fluences. We observe a strong decrease of the spectral weight for states below the Fermi level, as well as a filling of states above the Fermi level. In addition, the surface state $ss$ undergoes a slight shift toward higher binding energy. These changes occur on the 100 fs time scale and are evidently due to the photoexcitation induced by the pump pulse. 
Following excitation, the population of excited electrons relaxes on a 5-6 ps time scale because of electron-phonon coupling. This can be clearly seen in fig. \ref{fig_dynam}a which shows the evolution of excited electrons above the Fermi level (in the 50-200 meV range) as a function of time for a fluence of $0.84\;\mathrm{mJ/cm}^2$. While the initial filling of states above Fermi can be explained by electron excitation followed by electron thermalization, the decrease in the number of excited electrons is due to the cooling of the electron temperature by electron phonon-coupling. Clearly, fig. \ref{fig_dynam}a illustrates that electron thermalization and electron-phonon coupling occur on different time scales in bismuth.

Figure \ref{fig_spectra} also shows that at higher fluence, the number of excited electrons is of course larger. In addition, the comparison of panel a) and b) reveals a striking effect: the population of the empty surface state $ss'$ fills up faster at high fluence than at low fluence. To make this clearer, we have represented the amount of excited electrons in the 50-200 meV  range (corresponding to state $ss'$) as a function of time and for different fluences, see fig. \ref{fig_dynam}b. The number of excited electrons in this band clearly has a dynamics which depends on fluence and which can take longer than the pump pulse duration. The experimental points of fig. \ref{fig_dynam}b were fitted using the following function (full curves in fig. \ref{fig_dynam}b)
\begin{equation}
f(t)=A\left( 1+B\times \mathrm{erf}\left(\frac{t-t_0}{\tau}\right)\right)
\end{equation}
where $\mathrm{erf}(t)$ is the error function; $A$, $B$, $t_0 $ and $\tau$ are the fitting parameters. This procedure permits to extract the characteristic time $\tau_{ss'}$ for populating empty surface states $ss'$. The error function was chosen empirically because it is well adapted for fitting the behavior of a system reaching a quasi-stationary state monotonously. The variation of $\tau_{ss'}$ as a function of pump fluence is shown in fig. \ref{fig_dynam}d: at low fluence, $\tau_{ss'}$ can be as large as 200 fs, whereas it falls below the experimental resolution of 60 fs for fluences greater than $0.6\,\mathrm{mJ/cm^2}$. 

In fig. \ref{fig_dynam}c, we have also plotted the band shift of the surface state $ss$ as a function of time. Here again, the dynamics of the band shift is fluence dependent. The characteristic time for the band shift $\tau_{shift}$ follows closely the behavior of $\tau_{ss'}$ and occurs on a time which is faster than the period of the fastest phonon mode in bismuth. This indicates that the band shift has a purely electronic nature as it is sensitive to the rearrangement of the electron density. 

The characteristic time $\tau_{ss'}$ is evidently the electron thermalization time, after which it is safe to consider that the electron bath can be described by a Fermi-Dirac distribution. This is supported by the fact that all electronic observables, such as band depletion, band filling and band shifts reach a metastable electronic state after time $\tau_{ss'}$. Those quantities then all relax on a picosecond time scale, as seen in fig. \ref{fig_dynam}a, because of electron-phonon coupling: the electron bath exchanges energy with the phonon bath until the lattice temperature equalizes with the electron temperature. This part of the dynamics is well described by the two-temperature model \cite{kaga57,allen87} and it has been observed in many time-resolved reflectivity experiments.\cite{hase98,bosc08}

\subsection{Discussion}
In the following, we will discuss electron thermalization as the results of electron-electron scattering mainly. While electron-phonon scattering can also, in principle, play a role in thermalization, we stress that it operates on a longer time scale in bismuth (typically several to tens of picoseconds as explained earlier). Although it might be efficient for randomizing the electron momentum, it is not efficient for thermalizing the electron population on a sub-picosecond time scale, namely because phonons have low energies in bismuth ($< 20$ meV).

The fact that electron thermalization occurs on a longer time scale for low excitation fluences can be explained qualitatively using the theory of Fermi liquids.\cite{pines66} Strictly speaking, this theory is only valid for an elementary excitation in a material at equilibrium, with a given temperature. Thus we will use these arguments with care in the present context because following photoexcitation, the system is far from equilibrium and the electronic temperature is not well defined until the electron bath thermalizes. In the theory of Fermi liquids, an elementary excitation has a lifetime $\tau_l$ with\cite{pines66}
\begin{equation}\label{eqFermiLiqu}
\frac{1}{\tau_l} = K\frac{(E-\mu)^2+(\pi k_BT_e)^2}{1+\exp{[(\mu-E)/k_BT_e]}} 
\end{equation}
where $\mu$ is the chemical potential. The constant $K$ is $K=m^{*3}/(16\pi^4\hbar^6)\times \tilde{W}$, where $m^*$ is the electron effective mass and $\tilde{W}$ is an averaged transition probability.
Electrons which are promoted to high energy states by the pump laser pulse at 1.6 eV typically have energies $(E-E_F)\gg k_BT_e$. Since these electrons have a large phase space available for scattering events, we assume that their energy relaxation is extremely fast. As the electron population changes through electron electron scattering (i.e. impact ionization and Auger recombination), the states close to the Fermi level become more populated. Therefore, the electron population with $(E-E_F)^2\ll (\pi k_BT_e)^2$ increases monotonously in time toward the equilibrium value of a thermalized electron gas. Here, we identify $T_e$  with the electronic temperature that the system reaches after electronic thermalization is complete. The dynamics is slower for elementary excitations close to the Fermi level as Pauli's exclusion principle restricts the phase space volume which is available for scattering events. In this case, the theory of Fermi liquids predicts that the lifetime of a quasiparticle close to the Fermi level scales as $1/\tau_l\propto (k_BT_e)^2$. The temperature scaling may be different for the case of bulk bismuth because the density of states at the Fermi level is very small. Thus, one expects that the filling of electronic states close to the Fermi level will follow the trend of the quasiparticle lifetime with a modified scaling law. This behavior is in qualitative agreement with the experimental results: a low fluence leads to a lower electron temperature and thus to a slower filling of the band $ss'$.

\begin{figure}[t]
\begin{center}
\includegraphics[width=8.5 cm]{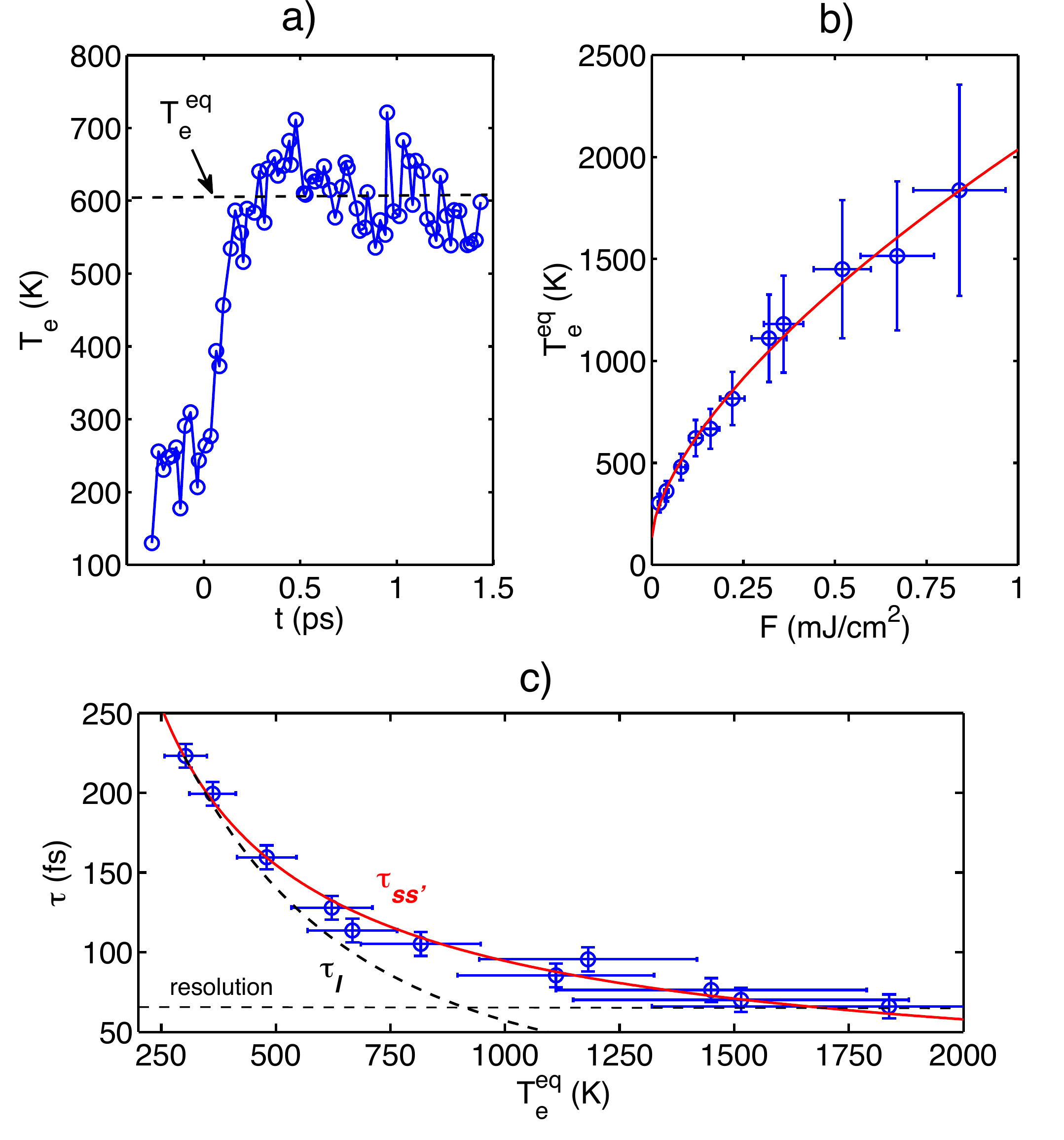}
\caption{(a) Evolution of the electronic temperature as a function of pump-probe delay for $F=0.16\;\mathrm{mJ/cm^2}$. (b) The thermalized electronic temperature as a function of laser pump fluence. The red curve is a fit $\propto F^\alpha$ with $\alpha=0.6$. (c) Electron thermalization time as a function of the thermalized electronic temperature $T_e^{eq}$. The full red curve is a fit $\propto 1/T^\beta$ with $\beta=0.7$. The dashed black curve represents $\tau_l$ computed using Eq.  \ref{eqFermiLiqu}.}
\label{fig_T}
\end{center}
\end{figure}

Note that this approach is only qualitative because the theory of Fermi liquids describes the relaxation of a single excited electron in a thermalized electron bath. This is not the case here: the whole electron population is initially out of equilibrium. Thus, the description of electron thermalization requires more sophisticated tools. Recently, Mueller and Rethfeld \cite{muel13} have developed a simplified approach in order to simulate electron relaxation in out of equilibrium conditions. Their model is based on the Boltzmann equation for the evolution of the electronic distribution function; it takes into account the density of states of a given material and uses a simplified description for the dispersion of electronic states. The relaxation time of the electronic system is defined thermodynamically: it is the time it takes for the entropy of the electronic sub-system to reach a maximum value. This approach shows that for simple metals and high electronic temperatures, the relaxation time is not too far from a $1/T_e^2$ dependence, as in Fermi liquids. However, the deviations from the Fermi liquid scaling turned out to be strongly material dependent. Moreover, the accuracy of the $1/T_e^2$ scaling in an interval of electronic temperatures comparable to our experimental conditions has not been discussed before. This motivated us to attempt to understand the behavior of the thermalization time of electrons in bismuth as a function of $T_e$. 

By monitoring the depopulation of the states below the Fermi level, it is possible to retrieve the value of the electron temperature (the exact procedure for computing the temperature is described in the Supplementary Information of Ref. \onlinecite{papa12}). Figure \ref{fig_T}a) shows the evolution of the electron temperature as a function of pump-probe delay for a incident pump fluence of $0.16\;\mathrm{mJ/cm^2}$. After a few hundred femtoseconds, the electron bath has thermalized and the electron temperature reaches an equilibrium value $T_e^{eq}$. The dependence of $T_e^{eq}$ with the pump fluence is shown in fig. \ref{fig_T}b: it can be fitted by a function $AF^{\alpha}$ with $\alpha=0.6$. Fig. \ref{fig_T}c) shows the dependence of the thermalization time with the equilibrium temperature. It can be fitted by a function $1/T^\beta$ with $\beta=0.7$. The black dashed curve represents the Fermi liquid behavior for electrons in state $ss'$ as computed using Eq. \ref{eqFermiLiqu}.  Clearly the experimental result deviates from a simple Fermi liquid. The slower thermalization time that we found experimentally could be due to the fact that bismuth is far from being a 3D degenerate electron gas, as the density of states is noticeably small at the Fermi level,\cite{xu96} and the band dispersion is strongly anisotropic.\cite{bene02,arn05} We also stress that a serious modeling of electron thermalization should also take into account dynamic screening. For instance, at the Bi (111) surface, the surface states account for a carrier density of $7\times10^{-3}$ electrons per bismuth atom in the low temperature ground state. Photoexcitation with a fluence of $1\;\mathrm{mJ/cm^2}$ generates a carrier density of $2\times10^{-2}$ electrons per Bi atom. This large change in the photoexcited carrier density could also have sizeable effects on the thermalization of electrons in Bi.

To conclude on electron thermalization, our photoemission data shows that the thermalization time at low fluence of $20\,\mathrm{\mu J/cm}^2$ is 225 fs, which is consistent with THz data of Ref. \onlinecite{iuri2012}, showing a typical time of 0.6 ps at $10\,\mathrm{\mu J/cm}^2$. Although the formation of electron pockets in local valleys of the conduction band was not observed directly in our experiment, our experimental results do not allow us to exclude this hypothesis.

\section{Coupling of electronic states to the $A_{1g}$ phonon mode}\label{sec_elph}

As the electrons are excited to high energy states and while they thermalize, the changes of the potential energy surface can launch atomic displacements through the excitation of coherent phonons. We now focus on this phenomenon which occurs in the first few picoseconds following photoexcitation.

\begin{figure}[h]
\begin{center}
\includegraphics[width=8.5 cm]{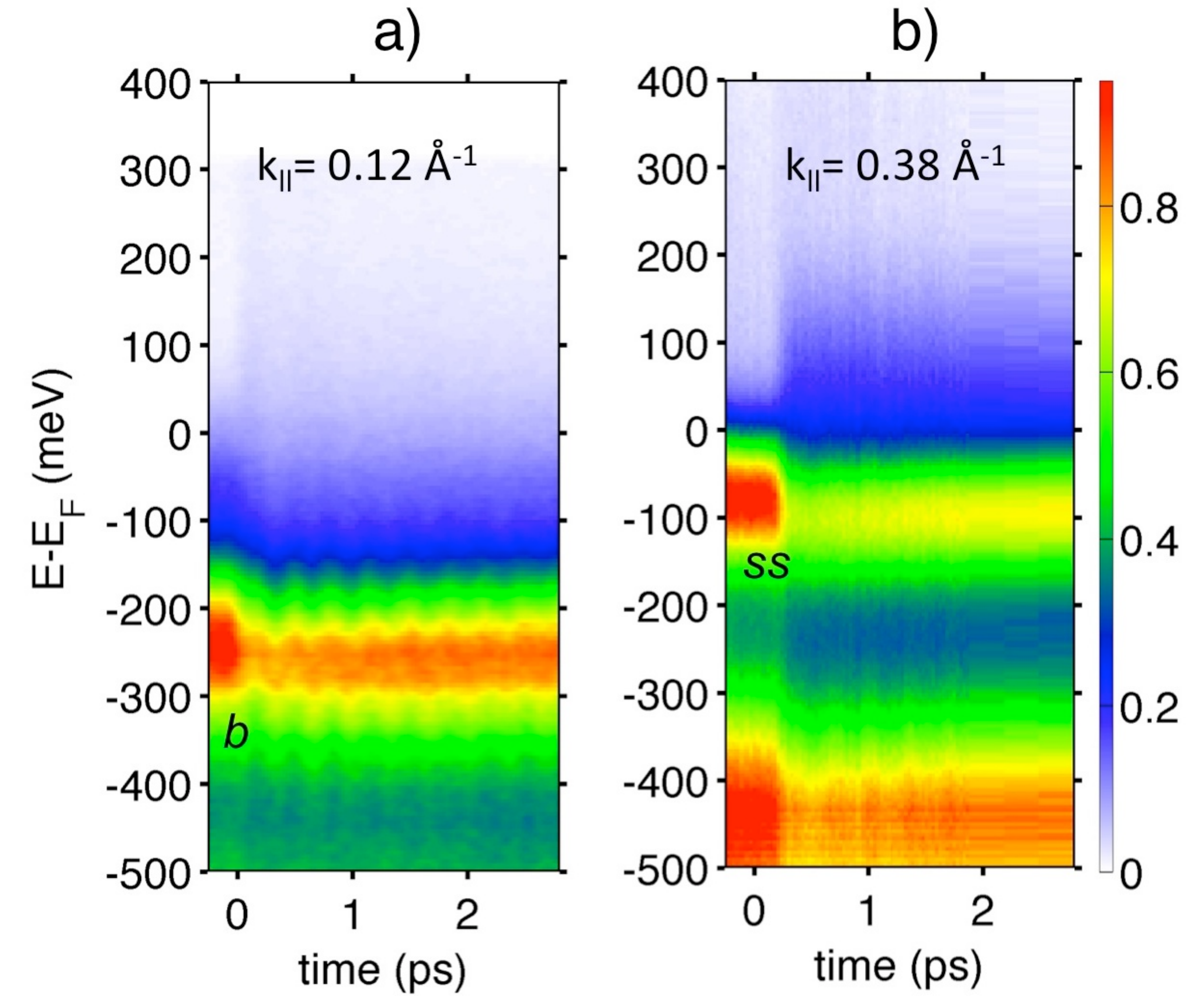}
\caption{a) Ultrafast dynamics of band \textit{b} following photoexcitation at $0.6\,\mathrm{mJ/cm^2}$. b) Ultrafast dynamics for the surface band \textit{ss}.}
\label{fig_dynam2}
\end{center}
\end{figure}

\begin{figure}[t]
\begin{center}
\includegraphics[width=8.5 cm]{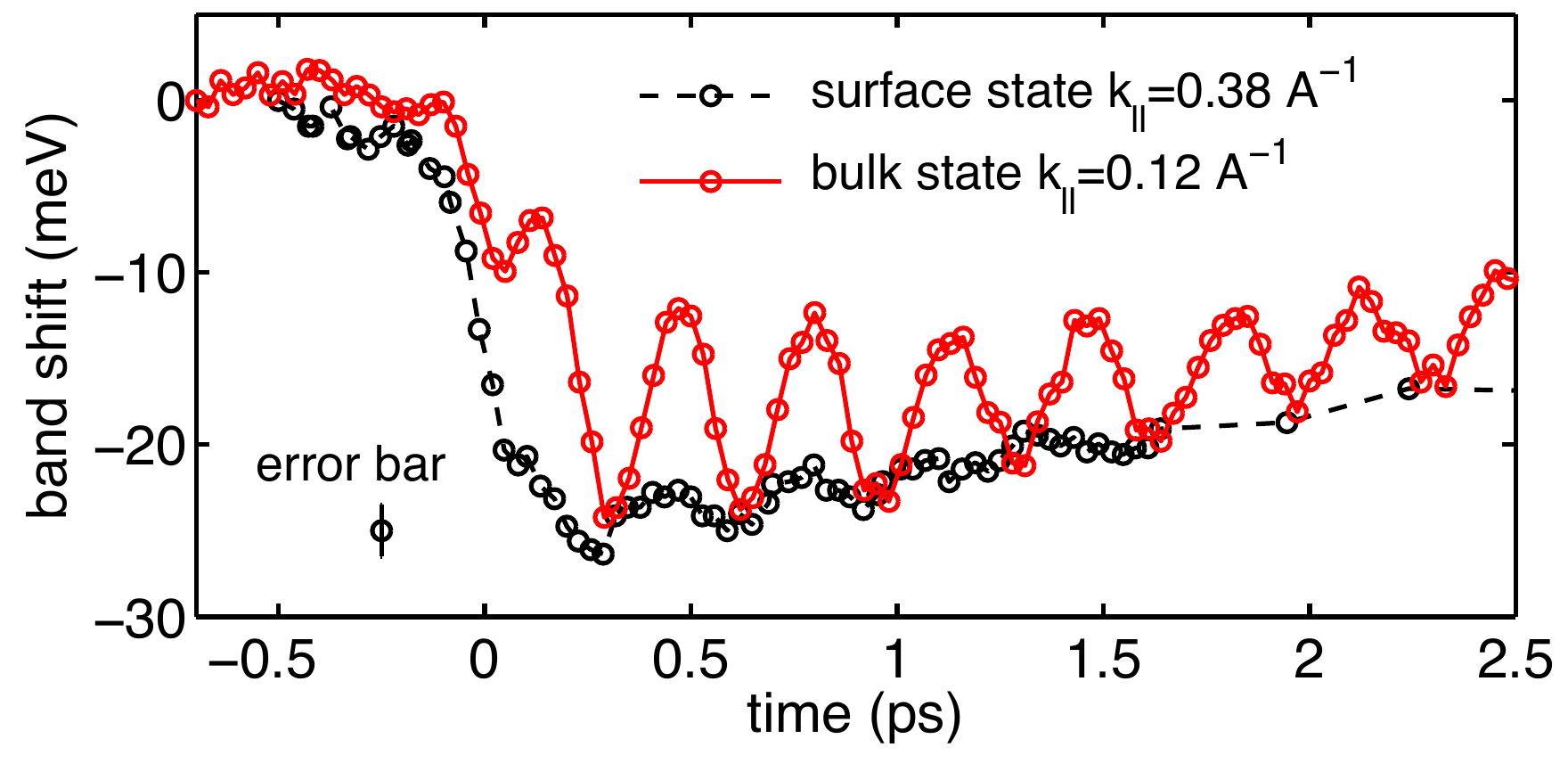}
\caption{Binding energy of the bulk state \textit{b} (full red line) and surface state \textit{s} (black dashed line) as a function of pump-probe delay. The error bar represents the typical energy error of the fitting procedure.}
\label{fig_scan}
\end{center}
\end{figure}

 Photoelectron spectra as a function of pump-probe delays are shown in fig. \ref{fig_dynam2}a for the bulk band at $k_\|=0.12\oneA$ and in fig. \ref{fig_dynam2}b for the surface band at $k_\|=0.38\oneA$. Again, we observe a strong decrease of the spectral weight for states below the Fermi level, as well as a filling of states above the Fermi level. The data was taken with a \textit{p}-polarized probe pulse for a better visualization of the states below $E_F$. In addition to the change of occupation numbers, we also observe modifications of the binding energies: both bands $b$ and $ss$ undergo a shift toward higher binding energy (as also seen in fig. \ref{fig_spectra}).  More importantly, band $b$ is clearly subject to large oscillations whereas band $ss$ does not seem to oscillate. To make this clearer, fig. \ref{fig_scan} represents the maxima of the bands as a function of pump probe delay. The bulk band clearly oscillates with an amplitude of about $15\pm2$ meV at 2.95 THz, which corresponds to the frequency of the $A_{1g}$ phonon mode. On the other hand, the surface band undergoes a very reduced oscillation: a weak oscillation of 3 meV amplitude can be observed but its significance is questionable as it is extremely close to the detection level (see the error bar on the figure). This is a general behavior of our data: bands with bulk character oscillate whereas surface states (such as $sr$ for instance) do not oscillate or have oscillation amplitudes close to the detection level. 


\begin{figure}[t]
\begin{center}
\includegraphics[width=8.5 cm]{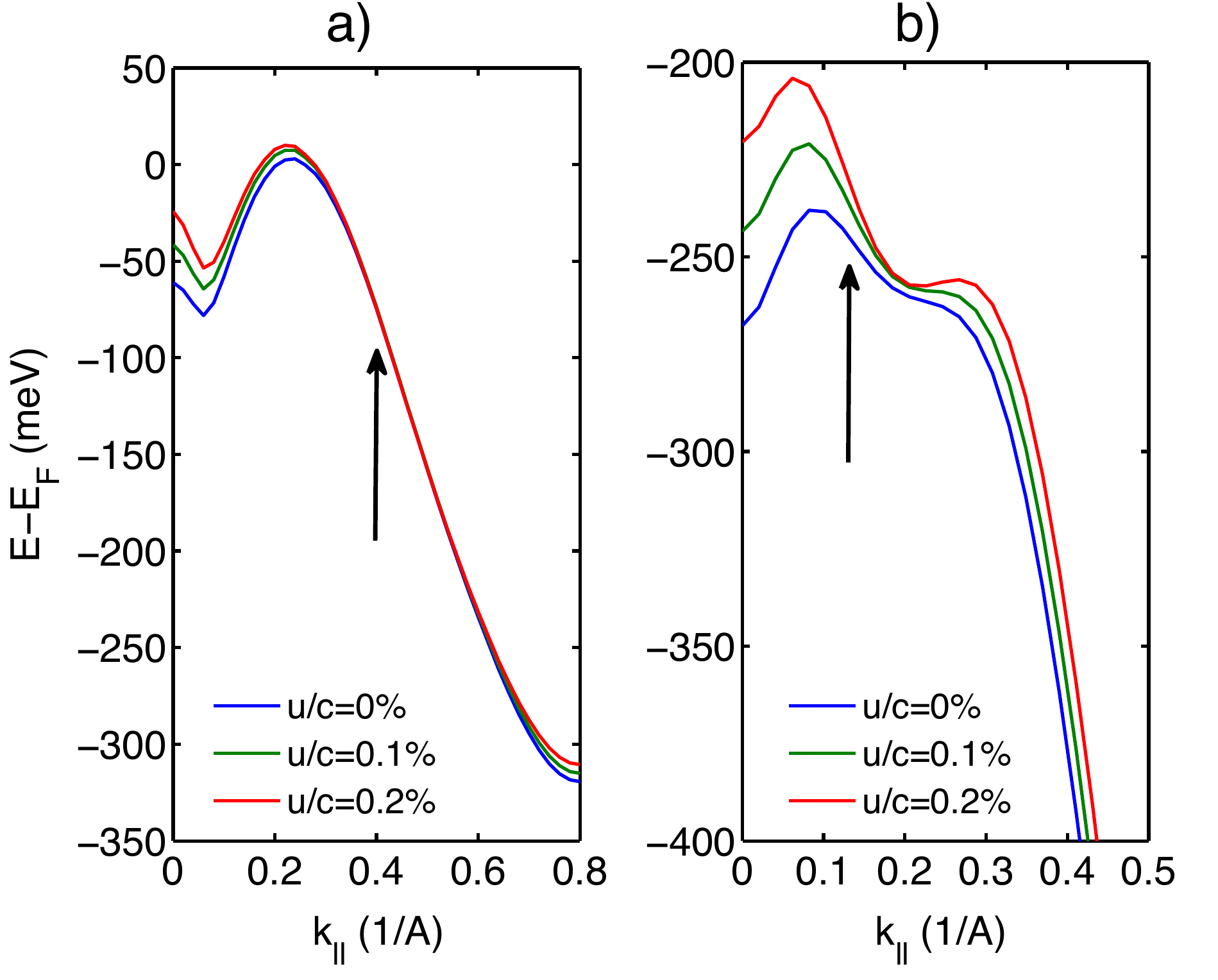}
\caption{DFT simulations of band dispersion in bismuth. The bands are calculated along the $\Gamma_s-\mathrm{M}_s$ direction and for various atomic displacements $u/c$ in the $A_{1g}$ mode. a) Dispersion of the surface state \textit{s} and b) bulk state \textit{b}.}
\label{fig_DFT2}
\end{center}
\end{figure}

The oscillations in spectral weight arise from the coupling of the $A_{1g}$ coherent phonon to electronic states. This effect can be assessed using DFT and calculations of the deformation potential. Let us considers that the two atoms in the bismuth unit cell are placed at $\pm(0.2329c+u)$, where $c=11.86\;\mathrm{\AA}$ is the length of the body diagonal and $u$ is the $A_{1g}$ phonon coordinate. We now consider a given energy state $|n,\mathbf{k} \rangle$ where $n$ is the band index and $\mathbf{k}$ the wave vector. Lattice motion in the $A_{1g}$ mode will shift the electronic level by $\delta E=D_n(\mathbf{k})u$. Furthermore, Allen\cite{allen84} has shown that in the case of an optical phonon, the deformation potential can be used to trace the matrix elements of the electron-phonon operator. For the $A_{1g}$ mode with $\mathbf{q}=0$, this translates into $\delta E=\langle n,\mathbf{k} |H_{el, ph}|n,\mathbf{k} \rangle$. Thus, our experiment gives us direct access to the electron-phonon coupling for various bands and various wave vectors.

While in previous work we have examined the wave vector dependence of electron-phonon coupling,\cite{papa12} we focus here on the dependence of electron-phonon coupling with the band index. The experimental results of fig. \ref{fig_dynam2} and \ref{fig_scan} imply that the electron-phonon matrix element for the bulk band is relatively large while it is small for the surface band. To confirm this, we have performed DFT simulations of both surface and bulk bands for various atomic displacement $u$. Dispersion of the bulk band $b$ for several values of $u$ are shown in fig. \ref{fig_DFT2}b. At $k_\|=0.12\,\mathrm{\AA}^{-1}$, the band shift $\delta E$ can reach 15 meV for displacement of $u/c=0.1-0.2\%$. Using a modified two temperature model,\cite{gire11,papa12,arn13} we found that the estimated atomic displacement is $u/c=0.14\%$ for an excitation fluence of $0.6\;\mathrm{mJ/cm}^2$. Consequently, according to fig. \ref{fig_DFT2}b, the calculated band shift is $16$ meV, in very good agreement with the experimental value of the oscillation amplitude. On the contrary, simulations of the surface band dispersion show that the energy is quite insensitive to atomic displacement in the $A_{1g}$ mode, see fig. \ref{fig_DFT2}a. In particular, at $k_\|=0.38\,\mathrm{\AA}^{-1}$, the band shift is smaller than 1 meV for $u/c=0.2\%$, i.e. below the experimental resolution. This weak coupling of surface states with the $A_{1g}$ mode explains why the surface band undergoes very reduced oscillations. These results also confirm that deformation potentials, as calculated in DFT, are a valid tool for estimating electron-phonon coupling, for both bulk and surface states.



\section{Conclusion}
In conclusion, we have investigated the ultrafast dynamics of the Bi (111) surface by means of time-resolved photoemission spectroscopy. Insights on electron thermalization could be obtained by measuring the filling of surface states. It was found that electron thermalization is strongly fluence dependent and can take place in hundreds of femtoseconds at low fluences, i.e. in a time longer than the laser pulse duration. The dependence on fluence could be explained qualitatively using scalings of quasi-particle lifetime obtained in the theory of Fermi liquids. However, a detailed analysis has revealed that the electronic thermalization time scales as $1/T_e^{0.7}$ instead of the $1/T_e^2$ dependence expected for individual quasiparticles in Fermi liquids. Finally, our photoemission data also gave us direct information on the coupling of electron states with the $A_{1g}$ phonon mode. We were able to show experimentally that surface states have a much weaker coupling to this mode when compared to bulk states, at least in the range of wave vectors where we performed our observations. These results were reproduced using DFT calculations and deformation potential calculations, showing that these numerical  tools can be trusted for evaluating the electron-phonon coupling as a function of wave vector and band index.

Finally, we also note that at low fluence, electron thermalization can take a relatively long time, up to several hundreds of femtoseconds, i.e. close to the period of the coherent phonon oscillation. This might complicate modeling since in the DECP mechanism and in the two temperature model, electron thermalization is assumed to occur within the duration of the pump pulse. Similarly, modifications of the potential energy surface are usually assumed to appear in a time shorter than the period of the coherent phonon. This approach is certainly valid at fluences higher that $1\,\mathrm{mJ/cm}^2$, that is in the domain where it was tested experimentally. \cite{fritz07,papa12} This might not be the case anymore at low fluences $<100\,\mathrm{\mu J/cm}^2$ where the potential energy surface is changing over several hundreds of femtoseconds.

\acknowledgments
The FemtoARPES project was financially supported by the RTRA Triangle de la Physique, the ANR program Chaires d'Excellence (Nr. ANR-08-CEXCEC8-011-01). We acknowledge partial financial support from EU/FP7 under the contract Go Fast, Grant agreement no. 280555. J. F. is supported by the European Research Council under Contract No. 306708.


\end{document}